\documentclass[10pt,prc,twocolumn,nofootinbib,preprintnumbers]{revtex4-2}
\usepackage{amssymb}
\usepackage{amsmath}
\usepackage{graphicx}
\usepackage[colorlinks=true, pdfstartview=FitV, linkcolor=blue, citecolor=blue, urlcolor=blue]{hyperref}

\begin{document}

\title{Impact of Effective Nucleon Mass and Multineutron States \\ on the Equation of State for Core-Collapse Supernovae }

\author{Tatsuya~Matsuki}
\affiliation{Department of Earth Science and Astronomy, The University of Tokyo, Tokyo 153-8902, Japan}

\author{Shun~Furusawa}
\affiliation{College of Science and Engineering, Kanto Gakuin University, Kanagawa, Japan}
\affiliation{Interdisciplinary Theoretical and Mathematical Sciences Program (iTHEMS), RIKEN, Wako, Saitama 351-0198, Japan}

\author{Kohsuke~Sumiyoshi}
\affiliation{National Institute of Technology, Numazu College, Shizuoka 410-8501, Japan}

\author{Hong~Shen}
\affiliation{School of Physics, Nankai University, Tianjin 300071, China}

\author{Katsuhiko~Suzuki}
\affiliation{Department of Physics, Tokyo University of Science, Tokyo 162-8601, Japan }

\begin{abstract}
In this study, we investigate the impact of effective nucleon mass and the existence of the dineutron~$(\mathrm{^{2}n})$ and the tetraneutron~$(\mathrm{^{4}n})$ on the thermodynamic properties and nuclear compositions by constructing new equations of state. 
Our results indicate that the model with a larger effective nucleon mass slightly alters the nuclear composition in neutron-rich environments primarily due to differences in the symmetry energy:
the mass fractions of unbound neutrons, protons, and heavy nuclei increase.
The impact on the thermodynamic properties is negligible, except for the chemical potentials.
On the other hand, multineutron states become prominent at high densities in neutron-rich environments, leading to a substantial reduction in the unbound neutron fraction. This depletion lowers the chemical potential of unbound neutrons, which in turn reduces the abundance of neutron-rich nuclei. Consequently, the number of unbound protons increases, leading to a corresponding rise in proton chemical potential.
These shifts in chemical potentials promote the formation of heavy nuclei with larger mass and atomic numbers. Ultimately, this compositional shift results in a lower free energy, primarily driven by the emergence of these heavy nuclei.
\end{abstract}

\maketitle

\section{Introduction}
 The nuclear equation of state~(EOS) is an essential input for core-collapse supernova 
 simulations, describing the pressure as a function of the baryon number density~$n_B$, the charge fraction~$Y_p$, and the temperature~$T$. 
The properties of uniform nuclear matter are explained within nuclear many-body theory, constrained by nuclear experiments and observational data from neutron stars~\cite{Oertel:2016bki,Furusawa:2022ktu}.
The relativistic mean-field theory~(RMF), one of the most commonly used approaches, describes nuclear forces as exchanges of $\sigma$, $\omega$, and $\rho$ mesons~\cite{sumiyoshi:1994ApJ,Shen:1998by,Shen:1998gq}.
Recently, Shen et al.\ constructed a new EOS based on the TM1e model, which introduces an $\omega$--$\rho$ interaction term to the TM1 model~\cite{Shen:2020sec}. This modification aims to refine the density dependence of the symmetry energy, thereby reproducing recent observational constraints on neutron star radii.
On the other hand, Li et al.\ constructed another model, the TM1m model, which reproduces the same saturation properties but with a larger effective nucleon mass, compared to the TM1e model~\cite{Li:2024tpr}. 
Variations in its value significantly impact the supernova dynamics such as the contraction of a proto-neutron star and the neutrino emission, at least based on the EOSs within the non-relativistic framework~\cite{Schneider:2019shi,Yasin:2018ckc}.

The EOS also determines the nuclear composition, which is a crucial factor in neutrino reactions. Non-uniform nuclear matter models are primarily classified into the single-nucleus approximation (SNA) and nuclear statistical equilibrium (NSE). 
The Thomas--Fermi (TF) approximation adopts the SNA by assuming that a representative nucleus forms a Wigner--Seitz cell~\cite{Shen:1998by,Shen_2011:,Togashi:2017mjp}. 
However, this approach tends to favor the formation of excessively heavy nuclei~\cite{Sumiyoshi:2022pqn}. 
In contrast, the NSE approach~\cite{Hempel:2009mc,Furusawa:2016tdj} accounts for the full distribution of nuclear species, as all nuclear reactions mediated by strong and electromagnetic interactions reach equilibrium at temperatures above approximately $0.4~\mathrm{MeV}$. 
Recent studies have demonstrated the necessity of incorporating detailed nuclear compositions, particularly light nuclei, which frequently appear in shock-heated regions~\cite{Ohnishi:2006mk,Furusawa:2013tta,Fischer:2015sll,Fischer:2020krf}. Furthermore, simulations comparing EOSs based on the SNA and NSE have shown that these compositional differences can influence the success of shock revival~\cite{Harada:2020fek} and the duration of neutrino emission from a proto-neutron star~\cite{Sumiyoshi:2022pqn}.

In neutron-rich environments, the possible implications of multineutron states – very weakly bound or resonant states consisting of neutrons – have attracted significant attention~\cite{Marques:2021mqf,Huang2023}. For example, recent nuclear experiments suggest that the dineutron~($\mathrm{^2n}$) and the tetraneutron~($\mathrm{^4n}$) can be formed within neutron-rich nuclei~\cite{Spyrou:2012zz,Duer:2022ehf}. Furthermore, the role of $\mathrm{^4n}$ in neutron stars and supernovae has been investigated~\cite{Ivanytskyi:2019ynz,Panov:2019khi,Pais:2023jdy}, revealing that these exotic states can emerge and influence nuclear compositions under neutron-rich conditions. 
In our previous study~\cite{Matsuki:2024ios}, we investigated the impact of $\mathrm{^2n}$ and $\mathrm{^4n}$ on the nuclear composition at 100 and 200~$\mathrm{ms}$ after core bounce, based on the results from a two-dimensional supernova simulation. We found that their presence increases the abundance of protons and deuterons, both of which play a vital role in neutrino reactions. However, that study did not account for the nuclear interactions of unbound nucleons or correlations among nuclei, which become increasingly significant at high densities where such exotic states can frequently emerge.

In the present study, we construct two types of EOSs within the extended NSE formalism, following the framework established by Hempel et al.~\cite{Hempel:2009mc}. 
First, we utilize the TM1e and TM1m models to describe the properties of uniform nuclear matter and unbound nucleons. 
We conduct a comparative analysis between the TM1e- and TM1m-based EOSs to examine the impact of the effective nucleon mass, while also comparing the TM1e-NSE EOS with the TM1e-TF EOS from a previous study~\cite{Shen:2020sec}. 
Hereafter, our TM1e-NSE and TM1m-NSE models are referred to as TM1e (NSE) and TM1m (NSE), respectively, while the model from the previous study is designated as TM1e (SNA). 
Second, we construct an EOS considering the existence of $^2\text{n}$ and $^4\text{n}$ based on the TM1e model. 
We investigate the influence of multineutron states by comparing the results of the TM1e~(NSE) with and without their inclusion. 
Hereafter, the NSE incorporating $^2\text{n}$ and $^4\text{n}$ is referred to as NSE-MN. 
In this paper, we use natural units with $\hbar = c = k_B = 1$.


{\section{Model}
We employ the extended NSE formalism based on the work by Hempel et al.~\cite{Hempel:2009mc}, with the following exceptions: the treatment of unbound nucleons, the definition of the volume fraction used for Coulomb energy shifts of heavy nuclei, and the nuclear excitation models for heavy nuclei.
The chemical potential of a nucleus with mass number $A$ and proton number $Z$ is given by
\begin{equation}
\label{eq:NSE}
\mu_{A,Z} = (A-Z)\mu_n + Z\mu_p,
\end{equation}
where $\mu_n$ and $\mu_p$ are the chemical potentials of an unbound neutron and proton, respectively. To minimize the free energy density, we impose the local conservation laws for the baryon number and the charge as follows:
\begin{equation}
    n_n+ n_p +\sum_{A,Z} n_{A,Z}A=n_B,
\end{equation}
\begin{equation}
    n_p + \sum_{A,Z}n_{A,Z}Z=n_BY_p,
\end{equation}
 where $n_n$, $n_p$ and $n_{A,Z}$ denote the number density of unbound neutrons, unbound protons, and nuclei, respectively.
\subsection{Unbound Nucleons}
We adopt the TM1e~\cite{Shen:2020sec} and TM1m~\cite{Li:2024tpr} models to describe the uniform nuclear matter above saturation densities and the nuclear interactions of unbound nucleons at subsaturation densities.
The Lagrangian of these models is given by
\begin{align}
\label{eq:Lagrangian}
\mathcal{L}
=& \sum_{i=p,n} \bar{\psi}_i
\Bigl[
 i\gamma_{\mu}\partial^{\mu}
 - \left(M + g_{\sigma}\sigma\right)
 \nonumber \\
&\qquad
 - \gamma_{\mu}
 \left(
   g_{\omega}\omega^{\mu}
   + \frac{g_{\rho}}{2}\tau_a\rho^{a\mu}
 \right)
\Bigr]\psi_i
\nonumber \\
&+ \frac{1}{2}\partial_{\mu}\sigma\,\partial^{\mu}\sigma
 - \frac{1}{2}m_{\sigma}^2\sigma^2
 - \frac{1}{3}g_{2}\sigma^{3}
 - \frac{1}{4}g_{3}\sigma^{4}
\nonumber \\
&- \frac{1}{4}W_{\mu\nu}W^{\mu\nu}
 + \frac{1}{2}m_{\omega}^2\omega_{\mu}\omega^{\mu}
 + \frac{1}{4}c_{3}
   \left(\omega_{\mu}\omega^{\mu}\right)^2
\nonumber \\
&- \frac{1}{4}R^a_{\mu\nu}R^{a\mu\nu}
 + \frac{1}{2}m_{\rho}^2\rho^a_{\mu}\rho^{a\mu}
\nonumber \\
&+ \Lambda_{\rm v}
 \left(g_{\omega}^2 \omega_{\mu}\omega^{\mu}\right)
 \left(g_{\rho}^2 \rho^a_{\mu}\rho^{a\mu}\right),
\end{align}
where $\psi$ denotes the nucleon field with the nucleon mass~$M=938~\mathrm{MeV}$, and $W^{\mu \nu}$ and $R^{a\mu \nu}$ represent the antisymmetric field tensors of $\omega$ and $\rho$, respectively. 
The coefficients $g_\sigma$, $g_\omega$ and $g_\rho$ denote the coupling constants of $\sigma$, $\omega$ and $\rho$ exchange.
In addition, the symbols $g_2$ and $g_3$ represent the coupling constants for the $\sigma$ self-interaction, while $c_3$ and $\Lambda_{\rm{v}}$ denote the strengths of the $\omega$ self-interaction and the $\omega$--$\rho$ coupling, respectively.
We apply the mean-field approximation, in which non-zero components are $\langle \sigma \rangle=\sigma$, $\langle \omega^0 \rangle=\omega$, and $\langle \rho^{30} \rangle=\rho$.
The detailed equations for the thermodynamic properties can be found in the previous study~\cite{Li:2024tpr}.

The coupling constants are determined by constraints on the saturation properties: the saturation density~$n_0$ and the energy per nucleon of symmetric nuclear matter~$E/A$, the incompressibility~$K$, the symmetry energy~$E_{\mathrm{sym}}$, and the symmetry energy slope~$L$ at saturation density. The TM1m model is constructed to reproduce the same saturation properties but with a larger effective nucleon mass,~$M^* =M+g_\sigma \sigma$. 
The coupling constants and saturation properties are shown in Tables~\ref{tab:CouplingConstant} and \ref{tab:saturation}, respectively. 
Since the density dependence of the symmetry energy in TM1m is weaker than in TM1e as illustrated in Fig.~3 of Ref.~\cite{Li:2024tpr}, the TM1m model predicts a smaller radius for a $1.4M_\odot$ neutron star. Specifically, the TM1m yields a radius of 12.4~km, whereas the TM1e predicts a larger radius of 13.1~km.

\begin{table*}[htbp]
  \centering
  \caption{Coupling constants of the TM1m and TM1e models~\cite{Li:2024tpr}.}
  \label{tab:CouplingConstant}
  \begin{tabular}{|c c c c c c c c|}
    \hline
    Model & $g_\sigma$ & $g_\omega$ & $g_\rho$ & $g_2 \; (\mathrm{fm}^{-1})$  & $g_3$  & $c_3$ & $\Lambda_{\mathrm{v}}$ \\
    \hline
    TM1m & 7.93528 & 8.63169 & 11.51296 & -11.51628 & 54.88715 & 0.00025 & 0.09326  \\
    TM1e & 10.0289 & 12.6139 & 13.9714 & -7.2325 & 0.6183 & 71.3075 & 0.0429   \\
    \hline
  \end{tabular}
\end{table*}


\begin{table*}[htbp]
  \centering
  \caption{Saturation properties in the TM1m and TM1e models~\cite{Li:2024tpr}.}
  \begin{tabular}{|c c c c c c c|}
    \hline
    Model & $n_0(\mathrm{fm^{-3}})$ & $E/A~\mathrm{(MeV)}$ & $K~\mathrm{(MeV)}$ & $E_{\rm{sym}}~\mathrm{(MeV)}$  & $L~\mathrm{(MeV)}$  & $M^*/M$\\
    \hline
    TM1m & 0.145 & -16.3 & 281 & 31.4 & 40 & 0.793  \\
    TM1e & 0.145 & -16.3 & 281 & 31.4 & 40 & 0.634  \\ 
    \hline
  \end{tabular}
  \label{tab:saturation}
\end{table*}

\subsection{Nuclei}
We incorporate a wide range of nuclear species based on experimental mass data~\cite{Wang:2021xhn} and theoretical predictions from the Finite-Range Droplet Model (FRDM)~\cite{Moller:2015fba}. To incorporate the excited-state data provided by Rauscher et al.~\cite{Rauscher:2003ti}, we primarily focus on the nuclei covered within their library. Specifically, we prioritize experimental mass data whenever available and utilize FRDM predictions to supplement unmeasured species.
Furthermore, we include certain species, such as light nuclei, which are absent from the Rauscher dataset if their experimental data are accessible. As a result, the nuclear species incorporated in this study are represented by the black and red shaded areas in Fig.~\ref{fig:Nucear_Map}.

Additionally, we incorporate $^2\mathrm{n}$ and $^4\mathrm{n}$. We adopt the typical binding energy values of $B_{^2\mathrm{n}} = -0.066$ MeV~\cite{Panov:2019khi} and $B_{^4\mathrm{n}} = -2.37$ MeV~\cite{Duer:2022ehf}, respectively. The binding energy~$B_{A,Z}$ is defined as $M_{A,Z} = (A-Z)m_n + Zm_p - B_{A,Z},$ where $M_{A,Z}$ denotes the mass of a nucleus.
Although the medium-density effects may modify their binding energies~\cite{Typel2010}, we utilize the theoretical and experimental values for simplicity.
In the NSE formalism, our previous paper demonstrated that their binding energies have a minor impact on the nuclear composition~\cite{Matsuki:2024ios}.

\begin{figure}[htbp]
  \centering
  \includegraphics[width=0.9\linewidth]{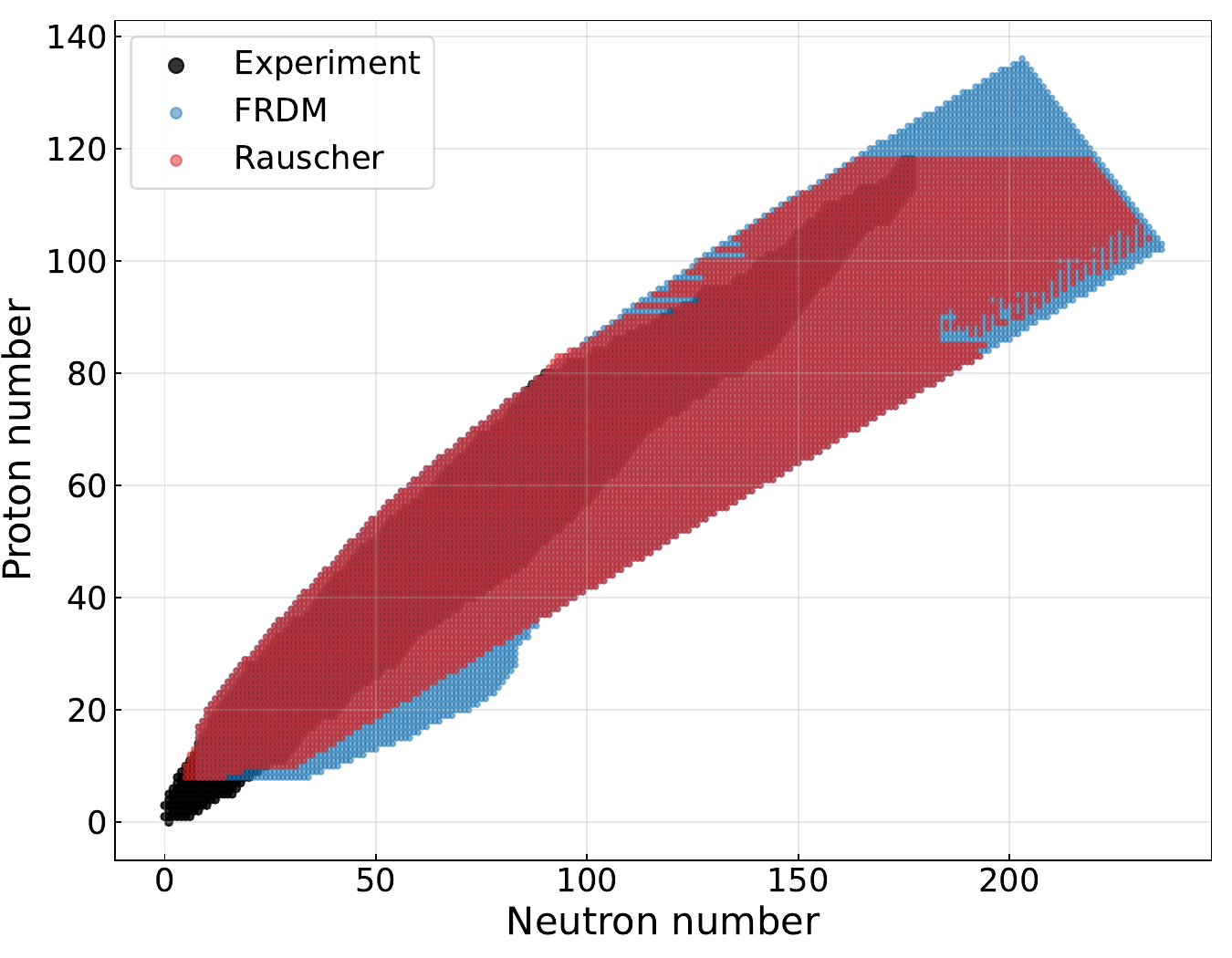}
  \caption{The distributions of experimental data~\cite{Wang:2021xhn}, FRDM~\cite{Moller:2015fba} and Rauscher data~\cite{Rauscher:2003ti}.}
  \label{fig:Nucear_Map}
\end{figure}

We also consider the Coulomb energy between nuclei and uniformly-distributed electrons, given by
\begin{equation}
\label{eq:Coulomb}
E_{\mathrm{Coul}} =
\frac{3}{5}\frac{Z^2 e^2}{R_N}
\left( -\frac{3}{2}x + \frac{1}{2}x^3 \right).
\end{equation}
Here, $e$ and $R_N$ represent the elementary charge and the radius of the nucleus, respectively. The latter is determined by the relation $(4\pi/3) R_N^3 n_0 = A$.
To avoid numerical complications, we assume the value of $x$ to be
\begin{equation}
x = \left(
\frac{n_B Y_p}{n_0}\frac{A}{Z}
\right)^{1/3}
\approx
\left(
\frac{n_B}{n_0}
\right)^{1/3}.
\end{equation}
This approximation is based on the expectation that in regions where heavy nuclei are dominant, the proton-to-mass ratio of the nuclei $Z/A$ tends to be comparable to the charge fraction of the system $Y_p$.

Moreover, we consider the excluded volume effect for unbound nucleons and nuclei. 
In contrast to the total volume $V$, the volumes available for nucleons~$V'$ and nuclei~$\bar{V}$ are reduced by nuclei and other baryons, respectively. 
The corresponding volume fractions for nucleons and nuclei are given by
\begin{equation}
\xi = \frac{V'}{V}
= 1 - \sum_{A,Z} \frac{n_{A,Z} A}{n_0},
\end{equation}
\begin{equation}
\kappa = \frac{\bar{V}}{V}
= 1 - \frac{n_B}{n_0},
\end{equation}
respectively.
We further define the local number densities as $n_n' = n_n / \xi$ and $n_p' = n_p / \xi$. 

In this formalism, the free energy density is given by
\begin{equation}
\label{eq:freedensity}
\begin{aligned}
f =\;&
\xi f^0_{\rm nuc}(n'_n,n'_p,T)
+ \sum_{A,Z} f_{A,Z}^0(n_{A,Z},T) \\
&+ f_{\rm Coul}(\{n_{A,Z}\})
- T \sum_{A,Z} n_{A,Z} \ln(\kappa),
\end{aligned}
\end{equation}
where $f_{\rm{nuc}}^0$ denotes the free energy density of unbound nucleons and 
the last term arises from the excluded volume effect.
The Coulomb free energy density and the free energy density of non-interacting nuclei are given by
\begin{equation}
f_{\rm Coul}(\{ n_{A,Z} \})
= \sum_{A,Z} n_{A,Z} E_{\mathrm{Coul}},
\end{equation}
\begin{equation}
\begin{aligned}
&f^0_{A,Z}(n_{A,Z},T)
 \\ = &n_{A,Z}
\Biggl[
M_{A,Z} 
- T \ln \left(
\frac{g_{A,Z}(T)}{n_{A,Z}}
\left(
\frac{M_{A,Z} T}{2\pi}
\right)^{3/2}
\right)
\Biggr],
\end{aligned}
\end{equation}
respectively, where $g_{A,Z}(T)$ denotes the internal degrees of freedom of nuclei, including excited states for heavy nuclei~\cite{Rauscher:2003ti}. 
In the EOS of Hempel et al., a simplified formula is employed, which tends to underestimate $g_{A,Z}(T)$ for heavy nuclei, as pointed out by Furusawa~\cite{Furusawa:2018xqk}.

In this case, the number density of nuclei is given by
\begin{equation}
\label{eq:nuclei}
\begin{aligned}
&n_{A,Z}
= \kappa
g_{A,Z}(T)
\left(
\frac{M_{A,Z} T}{2\pi}
\right)^{3/2} \\
&\quad \times
\exp\left(
\frac{
\mu_{A,Z}
- M_{A,Z}
- E_{\mathrm{Coul}}
- p_{\mathrm{nuc}}^0A/n_0
}{T}
\right),
\end{aligned}
\end{equation}
 where $p_{\mathrm{nuc}}^0$ denotes the pressure of unbound nucleons, given by Eq.~(4) in Ref.~\cite{Li:2024tpr}.

 When the density of the nucleon gas is sufficiently low ($n_n' + n_p' < 10^{-5}~\mathrm{fm}^{-3}$), we assume that the unbound nucleons behave as ideal gases, for simplicity.
In addition, when the unbound protons can be neglected ( $n_p'/(n_n'+n_p') <10^{-7} $), we assume the unbound nucleons form pure neutron matter.

\section{Results}
We calculate thermodynamic quantities and nuclear compositions  for proton fractions of $Y_p = 0.1, 0.3,$ and $0.5$, and temperatures of $T = 1, 5,$ and $10$ MeV to investigate the dependence on isospin asymmetry and temperature.

\subsection{Impact of the EOS models}
First, we compare the results of different models based on the TM1e~(NSE), TM1m~(NSE), and TM1e~(SNA).
Figure~\ref{fig:Mass_Fraction} shows the mass fraction $(X_{A,Z}=n_{A,Z}A/n_B)$ of unbound neutrons, protons, light nuclei~$(Z\leq5)$, and heavy nuclei~$(Z\geq6)$.
We find that the difference between TM1m (NSE) and TM1e (NSE) is negligible at $Y_p=0.3$ and $0.5$, while the difference becomes significant at $Y_p=0.1$ and in the region of high densities exceeding $10^{-2}\,\mathrm{fm^{-3}}$.
In this region, the number of neutrons and heavy nuclei in TM1m becomes slightly larger.

Figure~\ref{fig:composition} shows the detailed mass fractions of unbound nucleons and specific light nuclei. As observed in the figure, the abundances of protons and deuterons in TM1m~(NSE) increase at $Y_p=0.1$ and $T=5$ or $10$~MeV, compared to those in TM1e~(NSE). This behavior originates from the difference in the symmetry energy. 
The TM1m model exhibits a weaker density dependence of the symmetry energy, leading to a smaller difference in the chemical potentials, $\mu_n - \mu_p$. As a result, in the TM1m~(NSE) model, $\mu_n$ becomes lower (Fig.~\ref{fig:mun}) whereas $\mu_p$ becomes higher~(Fig.~\ref{fig:mup}), which causes the larger $X_p$ and $X_{\rm{^2H}}$. 
On the other hand, the weaker density dependence of the symmetry energy implies a reduced neutron repulsion, resulting in a larger $X_n$ in TM1m~(NSE).



These changes result in a slight increase in the average mass number and proton number of the heavy nuclei, as shown in Figs.~\ref{fig:A} and \ref{fig:Z}. These quantities are defined as
\begin{equation}
\langle A \rangle = \frac{\sum_{A,Z\geq 6} n_{A,Z} A}{\sum_{A,Z\geq 6}n_{A,Z}},
\end{equation}
and
\begin{equation}
\langle Z \rangle = \frac{\sum_{A,Z\geq 6} n_{A,Z} Z}{\sum_{A,Z\geq 6}n_{A,Z}},
\end{equation}
respectively.

On the other hand, significant discrepancies emerge between TM1e (NSE) and TM1e (SNA) at high densities and low charge fractions. Specifically, the TF approximation yields a larger mass fraction of heavy nuclei and higher average mass and proton numbers of heavy nuclei than the NSE model~(Figs.~\ref{fig:A} and~\ref{fig:Z}). This difference originates from the SNA within the TF framework, which tends to form the excessively heavy nucleus.

\begin{figure*}[htbp]
  \centering
  \includegraphics[width=\linewidth]{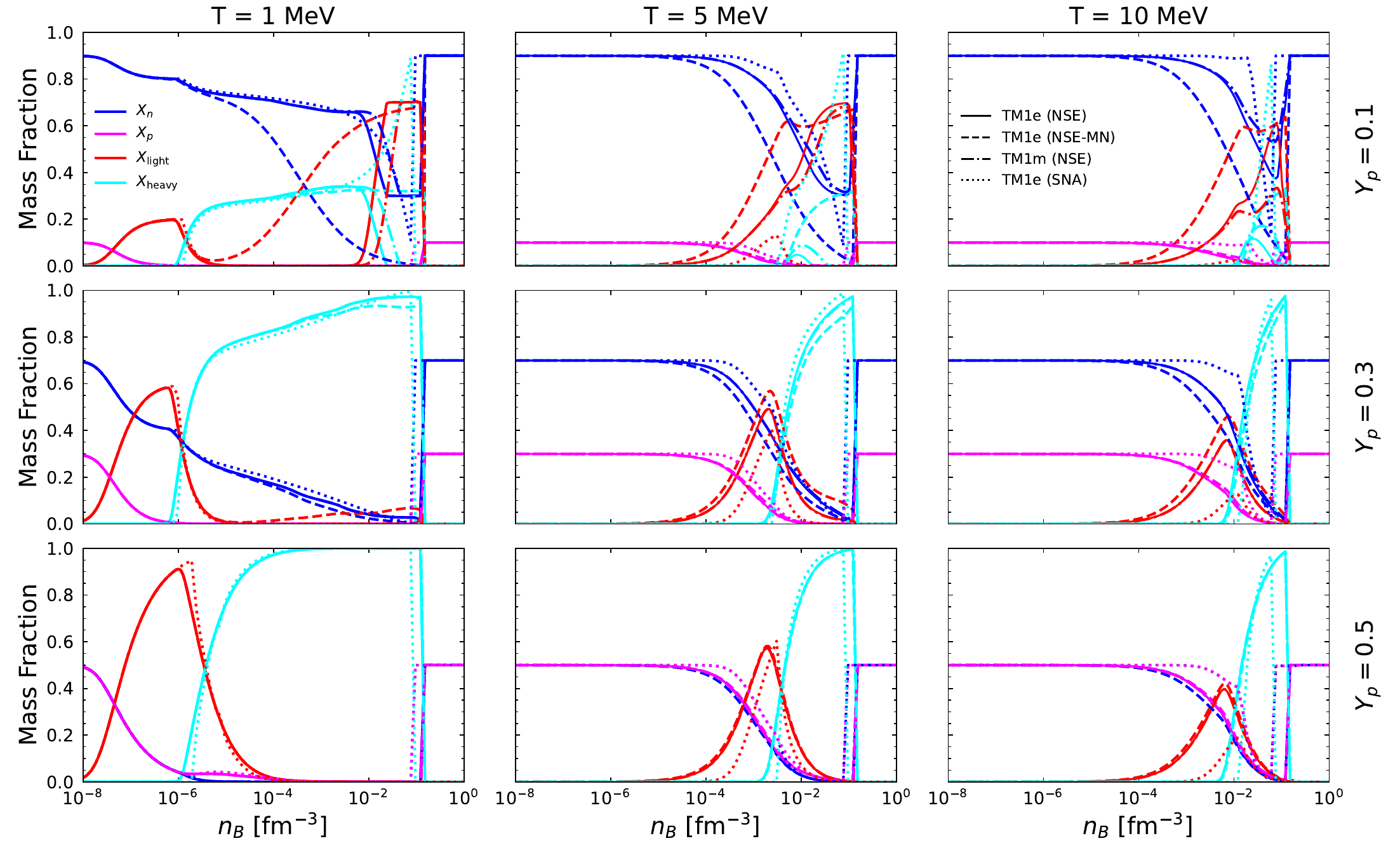}
  \caption{Mass fractions of unbound neutrons (blue), protons (magenta), light nuclei (red), and heavy nuclei (cyan).
  The solid, dashed, dashdot and dotted lines display the results of TM1e(NSE), TM1e(NSE-MN), TM1m(NSE) and TM1e(SNA), respectively. }
  \label{fig:Mass_Fraction}
\end{figure*}

\begin{figure*}[htbp]
  \centering
  \includegraphics[width=\linewidth]{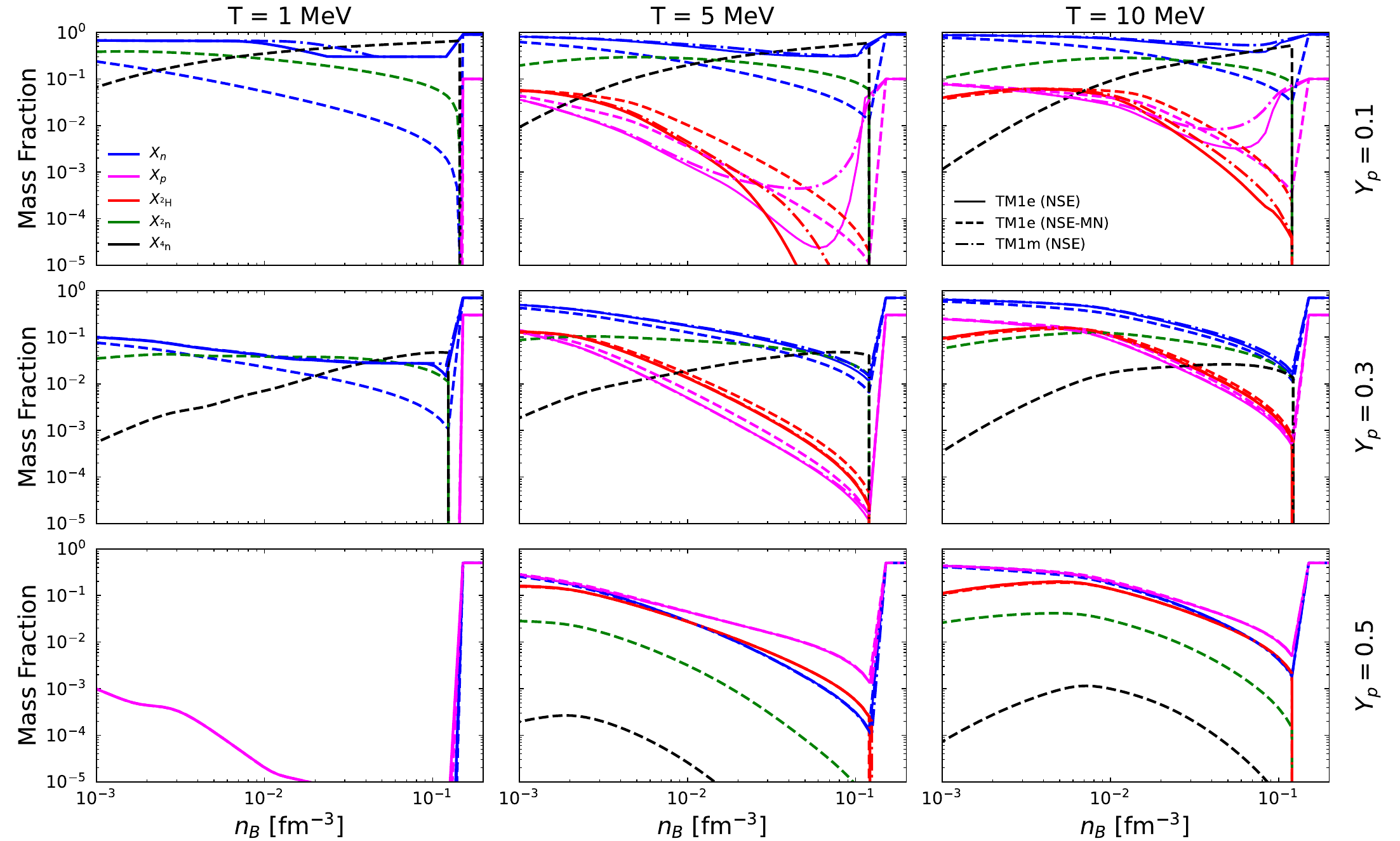}
  \caption{Detailed mass fractions of unbound neutrons~(blue), protons~(magenta), deuterons~(red), $\mathrm{^2n}$~(green), and $\mathrm{^4n}$~(black). Line styles are the same as in Fig.~\ref{fig:Mass_Fraction}, without the TM1e~(SNA) model.}
  \label{fig:composition}
\end{figure*}

\begin{figure*}[htbp]
  \centering
  \includegraphics[width=\linewidth]{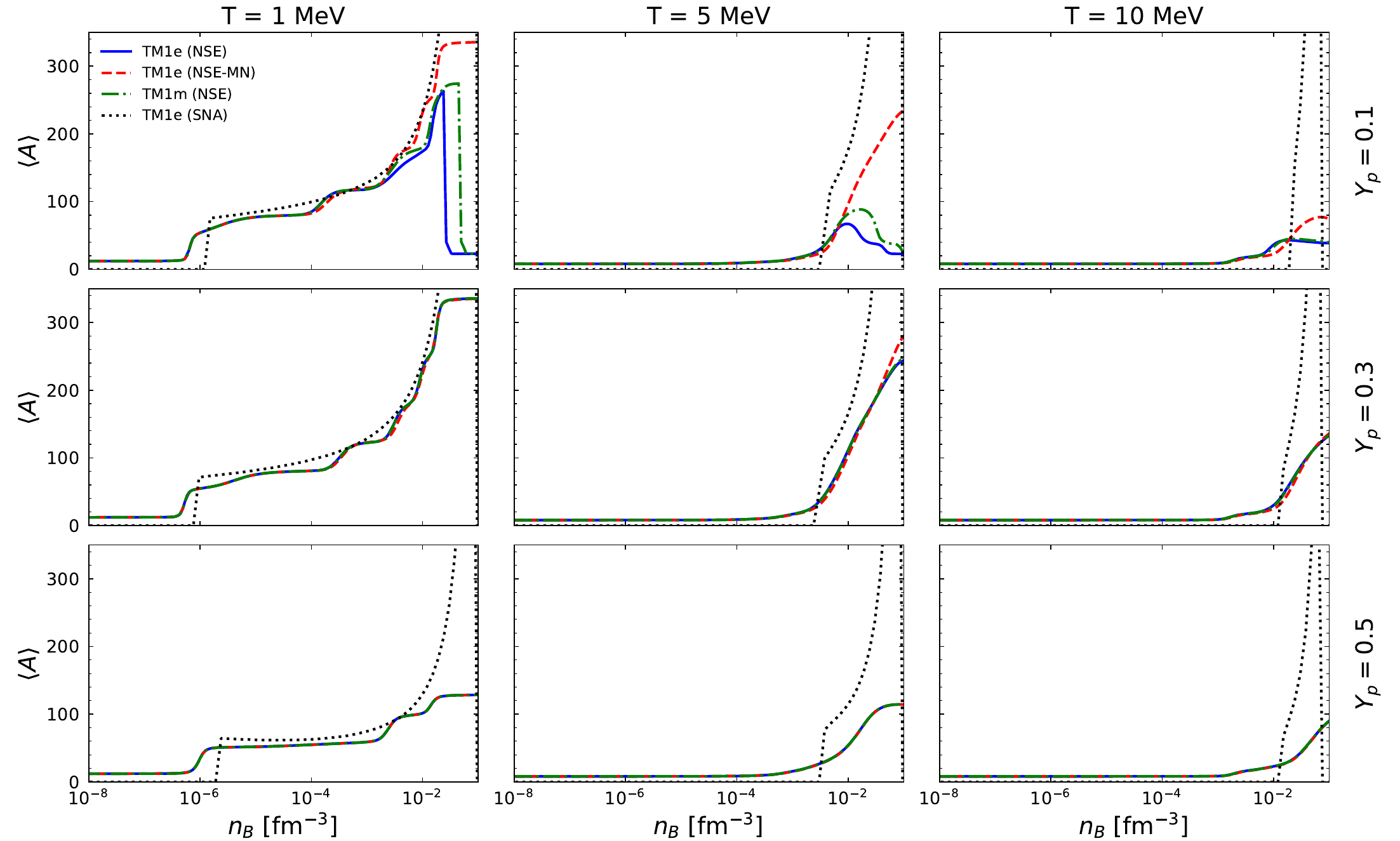}
  \caption{The average mass number of heavy nuclei~$\langle A\rangle$. 
  Line styles are the same as in Fig.~\ref{fig:Mass_Fraction}.}
  \label{fig:A}
\end{figure*}

\begin{figure*}[htbp]
  \centering
  \includegraphics[width=\linewidth]{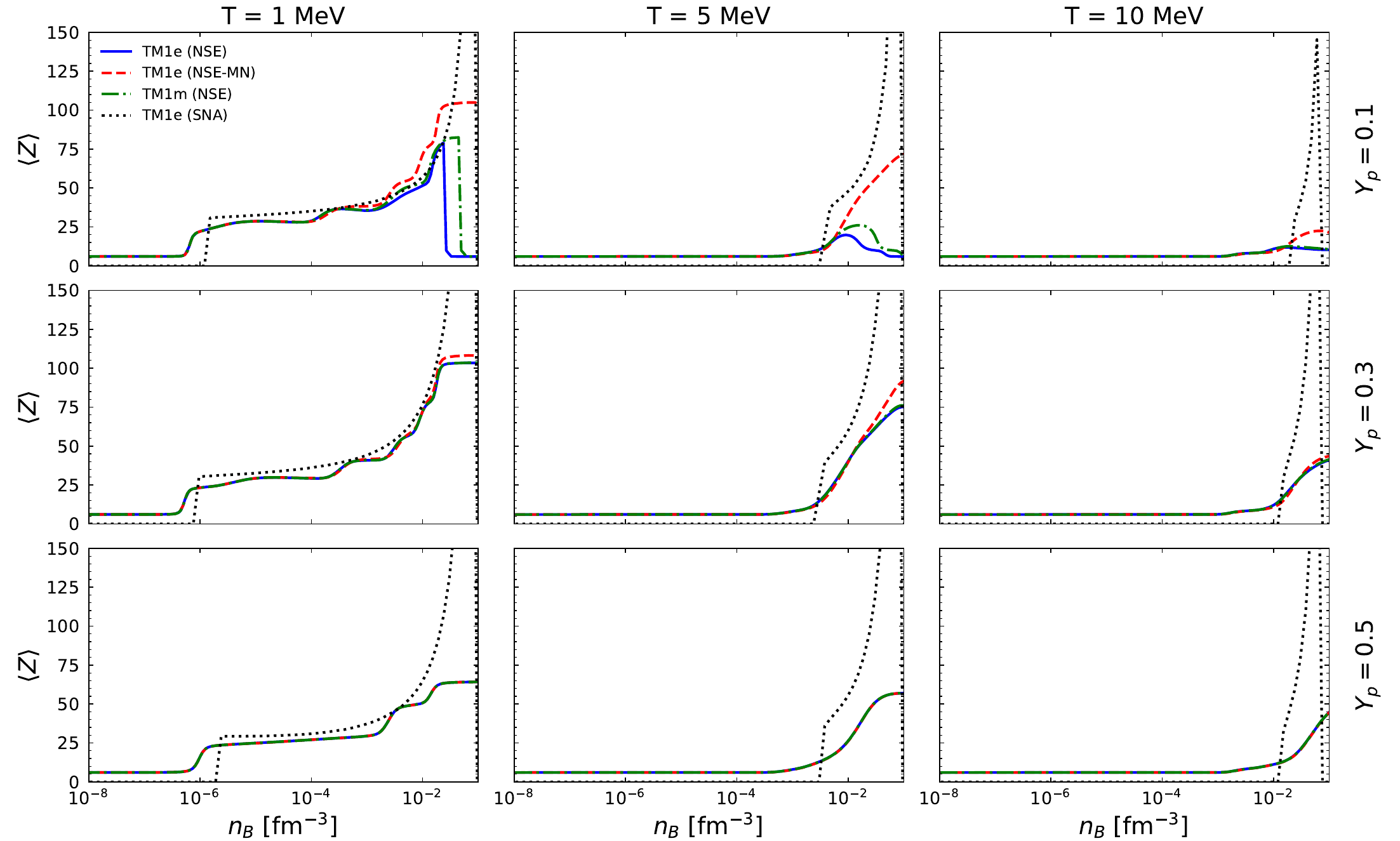}
  \caption{The average proton number of heavy nuclei~$\langle Z\rangle$. 
  Line styles are the same as in Fig.~\ref{fig:Mass_Fraction}.}
  \label{fig:Z}
\end{figure*}

Next, we proceed to the thermodynamic quantities. As observed in the following figures, the differences in the thermodynamic quantities between TM1e (NSE) and TM1m (NSE) are negligible, with the exception of the chemical potentials as mentioned above. Therefore, the subsequent discussion focuses primarily on the difference between the NSE formalism and the TF approximation. 

Figures~\ref{fig:mun} and \ref{fig:mup} display the chemical potentials of unbound neutrons and protons, respectively. Discrepancies between the NSE and TF results emerge near the saturation density, particularly at $T=10$~MeV with $Y_p=0.3$ or $0.5$. This reflects the different treatments of heavy nuclei formation and their impact on the surrounding nucleon gas. 

Figure~\ref{fig:Free} shows the free energy per baryon $F=f/n_B-m_u$, where $m_u$ denotes the atomic mass unit.
At $T=1$ MeV, small but consistent discrepancies appear at low densities. 
These differences arise because the NSE formalism incorporates experimental and theoretical data for nuclear masses, whereas the TF approximation does not explicitly account for individual mass data. Consequently, the pairing and shell effects are absent in the TF framework.
At this temperature, the dependence on the non-uniform nuclear matter model becomes significant near $10^{-2} \text{ fm}^{-3}$ and at $Y_p=0.3$ and $0.5$. This is because the additional degrees of freedom provided by the coexistence of various nuclear species contribute to maximizing the entropy and minimizing the free energy, as illustrated in Figs.~\ref{fig:Free} and \ref{fig:Entropy}.

At a higher temperature of $T=10$ MeV and charge fraction of $Y_p = 0.3$ and $0.5$, these discrepancies become more pronounced.  The model dependence stems from a significant decrease in the internal energy within the NSE framework (Fig.~\ref{fig:Eint}). 
By incorporating greater statistical degrees of freedom through a wide range of nuclear species, the NSE approach identifies a more energetically favorable composition compared to the TF approximation. 

Figure~\ref{fig:Pressure} shows the baryonic pressure. 
A major distinction between the NSE and TF results is the presence or absence of a pressure drop under certain conditions (e.g., $Y_p = 0.3$ and $T = 5$ MeV). This behavior is primarily attributed to the Coulomb interactions among nuclei. 
The nuclei with smaller mass numbers and proton numbers predicted in the NSE formalism lower the Coulomb attraction.
Furthermore, the SNA employed in the TF model neglects the translational motion of heavy nuclei, which leads to an underestimation of the positive thermal pressure. 

\subsection{Impact of multineutron states}
 We also investigate the impact of $\mathrm{^{2}n}$ and $\mathrm{^{4}n}$ 
 by comparing the results of TM1e~(NSE) and TM1e~(NSE-MN) models. 
As shown in Fig.~\ref{fig:Mass_Fraction}, the inclusion of these states significantly alters the nuclear composition, particularly at low charge fractions. At $Y_p = 0.1$, the abundance of unbound neutrons decreases substantially, while the mass fractions of light and heavy nuclei increase. 
The reduction in $X_n$ leads to a decrease in the chemical potential of unbound neutrons~(Fig.~\ref{fig:mun}). This results in a decrease in the abundance of neutron-rich nuclei, leaving excess protons that would otherwise form neutron-rich nuclei. The larger abundance of unbound protons leads to an increase in the proton chemical potential~(Fig.~\ref{fig:mup}). These shifts in chemical potentials increase the abundance of heavy nuclei.

The series of changes at all temperatures is driven by the prominent emergence of $\mathrm{^{2}n}$ and $\mathrm{^{4}n}$ at densities around $10^{-3}\,\mathrm{fm}^{-3}$, as illustrated in Fig.~\ref{fig:composition}. 
Notably, the abundances of unbound protons and deuterons in TM1e~(NSE-MN) also increase in most of these density regions, compared to those in TM1e~(NSE). 
On the other hand, at $Y_p=0.1$ and $T=5$ or $10$~MeV, the proton abundance near the saturation density decreases in the model including multineutron states. This reduction is attributed to the excluded-volume effect; while the disappearance of heavy nuclei would typically increase the abundance of unbound nucleons, the appearance of $\mathrm{^{2}n}$ and $\mathrm{^{4}n}$ significantly suppresses this increase.

The presence of $\mathrm{^{2}n}$ and $\mathrm{^{4}n}$ also exerts a substantial influence on the average mass number and proton number, as shown in Figs.~\ref{fig:A} and \ref{fig:Z}. At $Y_p = 0.1$, both values increase when multineutron states are incorporated. 
 This is linked to the increase in the unbound proton chemical potential.


Regarding thermodynamic properties, the appearance of multineutron states reduces the free energy at $Y_p = 0.1$ (Fig.~\ref{fig:Free}). 
This results from the formation of heavy nuclei, which decreases the internal energy, as seen in Fig.~\ref{fig:Eint}. In particular,
at $T = 1$~MeV, the increase in entropy is observed in Fig.~\ref{fig:Entropy}, due to an increase in light nuclei.
In contrast, the significant emergence of multineutron states leads to a reduction in pressure at $Y_p = 0.1$ and $T = 1$~MeV (Fig.~\ref{fig:Pressure}). This reduction primarily stems from the decrease in the number density of unbound neutrons. 

\section{Summary and Discussion} 
In this study, we developed new EOSs based on an extended NSE formalism and the TM1e and TM1m models, both of which describe the underlying uniform nuclear matter and nucleon interactions at subsaturation densities.
We have found that the choice of nuclear interaction model has only a minor influence on the thermodynamic properties, whereas it affects the composition in neutron-rich environments.
Compared to the results of TM1e (NSE), the weaker density dependence of the symmetry energy in the TM1m model predicts a higher abundance of unbound neutrons at high densities. Additionally, this feature leads to a decrease in the neutron chemical potential and a corresponding increase in the proton chemical potential. Consequently, TM1m (NSE) predicts a greater abundance of protons and heavy nuclei characterized by larger mass and proton numbers.

Moreover, the treatment of non-uniform nuclear matter highlights a fundamental difference between methods: while the TF approach tends to form a single representative heavy nucleus with an excessively large mass, the NSE formalism provides a realistic distribution of various nuclear species. These methodological differences result in distinct thermodynamic properties, most notably the characteristic pressure drop driven by Coulomb interactions. 

As new degrees of freedom, we have constructed a new EOS considering the existence of dineutrons and tetraneutrons.  
In neutron-rich matter, the formation of $\mathrm{^2n}$ and $\mathrm{^4n}$ significantly decreases $X_n$. This shift is accompanied by a decrease in the neutron chemical potential and a corresponding increase in the proton chemical potential. Such changes promote the abundance of unbound protons, deuterons, and heavy nuclei. 
On the other hand, a decrease in protons is observed at $Y_p = 0.1$ and $T = 5$ or $10$~MeV in the high-density region around $10^{-1}~\mathrm{fm^{-3}}$, where the excluded-volume effect becomes prominent.
In this regime, the substantial abundance of $\mathrm{^{2}n}$ and $\mathrm{^{4}n}$ effectively suppresses the proton abundance.
Overall, the resulting shifts in nuclear compositions due to multineutron states lead to a lower free energy at $Y_p=0.1$.

The differences in effective nucleon mass and the emergence of multineutron states affect the number of unbound protons at near-saturation densities and in neutron-rich environments, such as proto-neutron stars.
Since they play an essential role in neutrino emission, the larger effective mass and the emergence of multineutron states may increase neutrino emission from a proto-neutron star.
In addition, these factors increase the abundance of heavy nuclei with large mass number.
Since the coherent neutrino-nucleus scattering cross section is proportional to $A^2$, this enhancement in $\langle A \rangle$ could significantly strengthen neutrino scattering. Such an effect may lead to more efficient neutrino trapping and potentially extend the duration of neutrino emission compared to NSE-based models neglecting these states~\cite{Sumiyoshi:2022pqn}.

We are currently constructing EOS tables based on the present models for implementation in supernova and proto-neutron star cooling simulations. In future work, we will employ these simulations to systematically investigate how the effective nucleon mass and multineutron states influence the overall dynamical evolution.

In this study, several areas for further improvement remain.
First, the binding energies employed in this study are taken from the experimental values and theoretical predictions. However, in-medium modifications at high densities and finite-temperature effects could alter these binding energies, potentially impacting the predicted nuclear compositions.

Second, the treatment of multineutron states may not suit the standard assumptions of the NSE formalism. A more rigorous incorporation of these states may require the use of a robust formalism~\cite{Voskresenskaya:2012np, Shen_2011:,Ropke:2014fia,Typel2010}.

Third, while we utilized the excluded-volume method to describe the dissolution of nuclei during the transition to uniform nuclear matter, this approach has inherent limitations. Specifically, it does not account for the species-dependent dissolution behavior of individual nuclei~\cite{Pais:2023jdy}. For example, studies employing generalized density-functional approaches suggest that the excluded-volume method may overestimate nuclear abundances compared to models that explicitly treat density-dependent dissolution~\cite{Fischer:2020krf}.

Finally, our results rely on the partition functions from Rauscher et al.~\cite{Rauscher:2003ti} for the excited states. However, the choice of the excited-state model remains a significant source of uncertainty. As demonstrated by Furusawa et al.~\cite{Furusawa:2018xqk}, different treatments of finite-temperature effects in the NSE formalism can lead to appreciable variations in the resulting EOS. 
The integration of these factors remains for future work.

\begin{acknowledgments}
We thank K. Mameda for helpful discussions. S. Furusawa is supported by a Grant-in-Aid for Scientific Research (Grant No. 26K07073). K. Sumiyoshi is supported by Grants-in-Aid for Scientific Research (Grant Nos. 24K00632 and 25H01273). K. Suzuki is supported by a Grant-in-Aid for Scientific Research (Grant No. 25K07307). 
K. Sumiyoshi acknowledges 
Computing Research Center, KEK, 
JLDG on SINET of NII, 
Research Center for Nuclear Physics, Osaka University, 
Yukawa Institute of Theoretical Physics, Kyoto University, 
and 
Information Technology Center, University of Tokyo 
for providing high performance computing resources.
This work was partially supported by the computing resources of the Center for Computational Astrophysics, National Astronomical Observatory of Japan.
\end{acknowledgments}

\begin{figure*}[htbp]
  \centering
  \includegraphics[width=\linewidth]{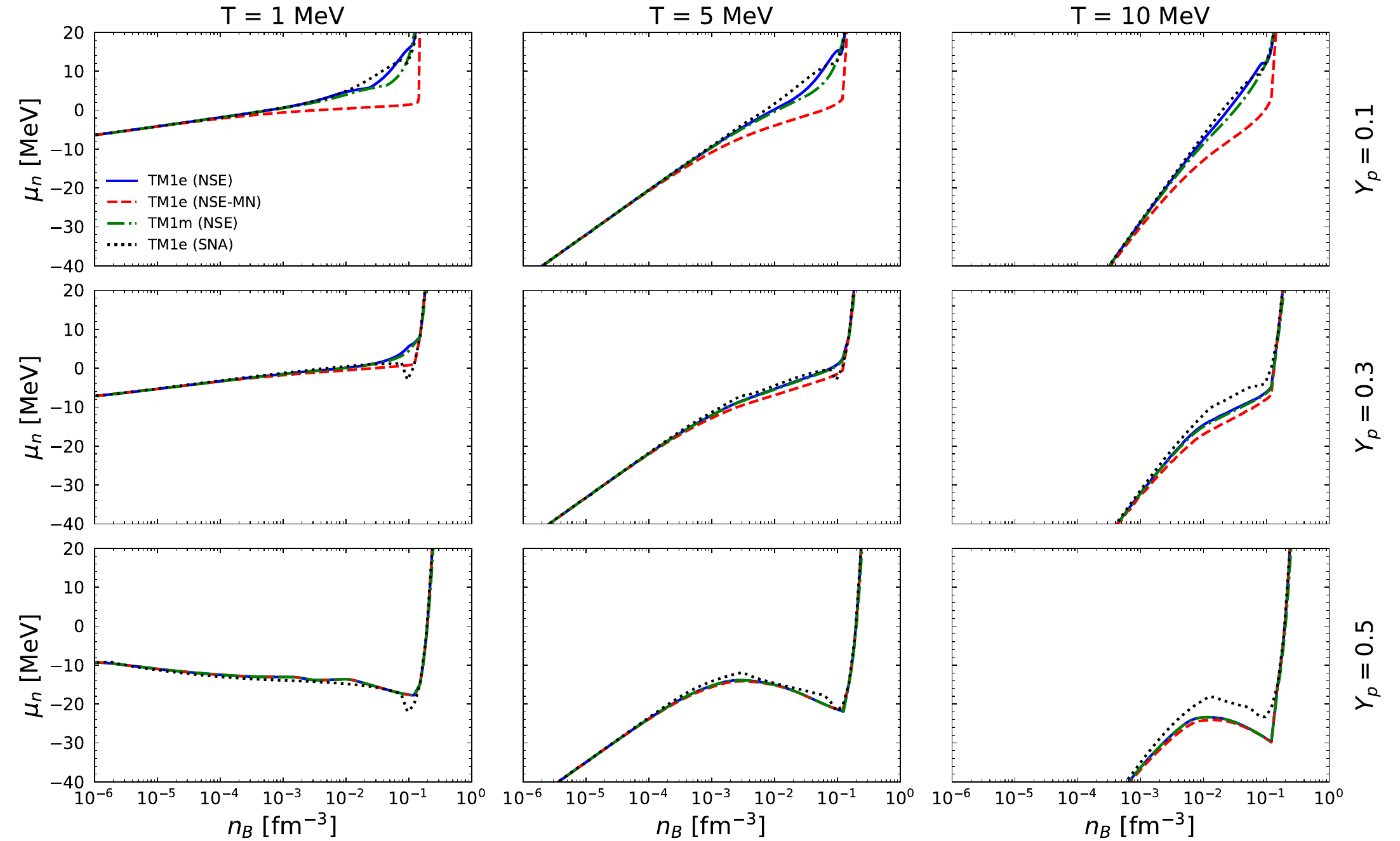}
  \caption{Chemical potential of unbound neutrons $\mu_n$.  Line styles are the same as in Fig.~\ref{fig:Mass_Fraction}.}
  \label{fig:mun}
\end{figure*}

\begin{figure*}[htbp]
  \centering
  \includegraphics[width=\linewidth]{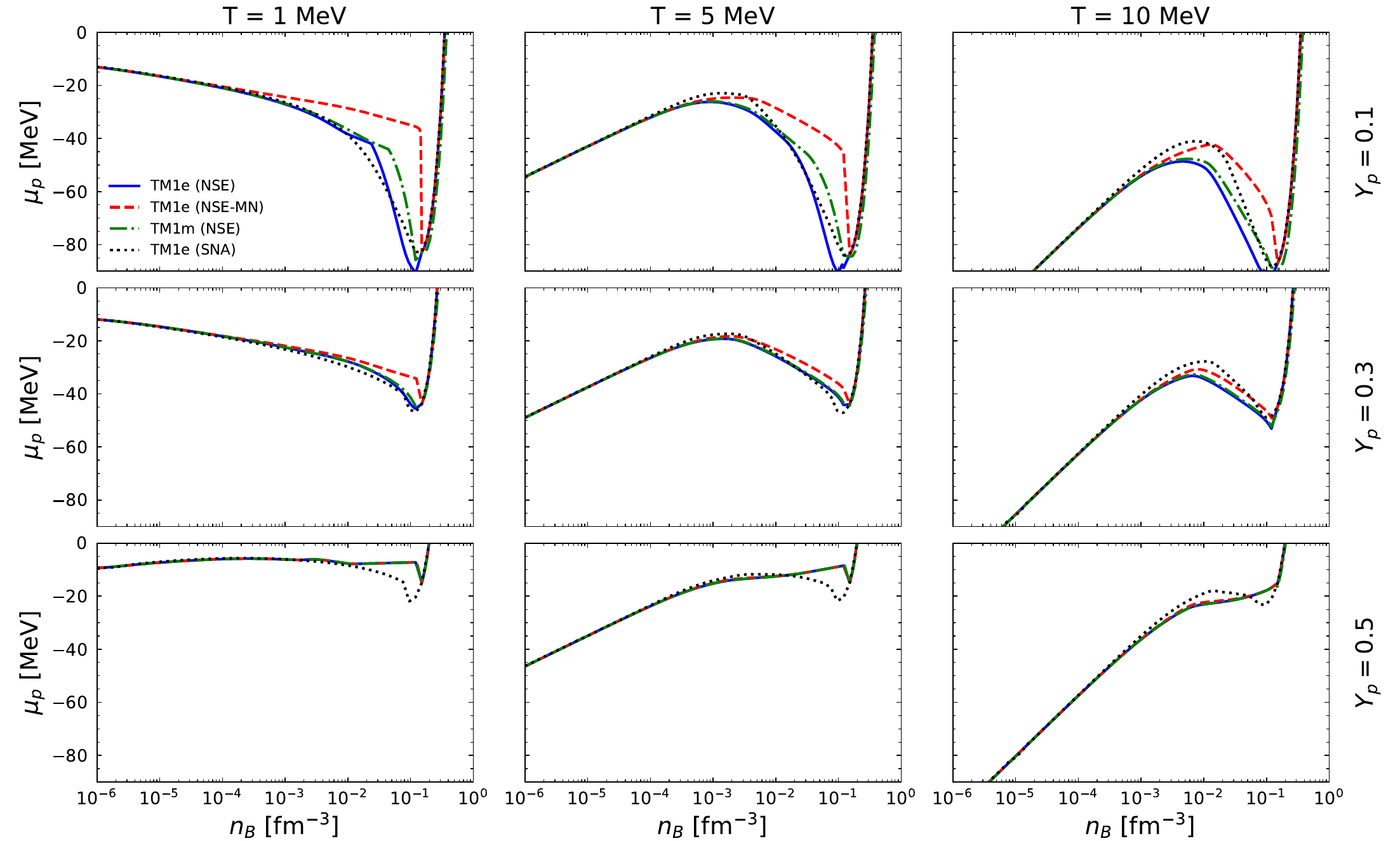}
  \caption{Chemical potential of unbound protons $\mu_p$. Line styles are the same as in Fig.~\ref{fig:Mass_Fraction}.}
  \label{fig:mup}
\end{figure*}

\begin{figure*}[htbp]
  \centering
  \includegraphics[width=\linewidth]{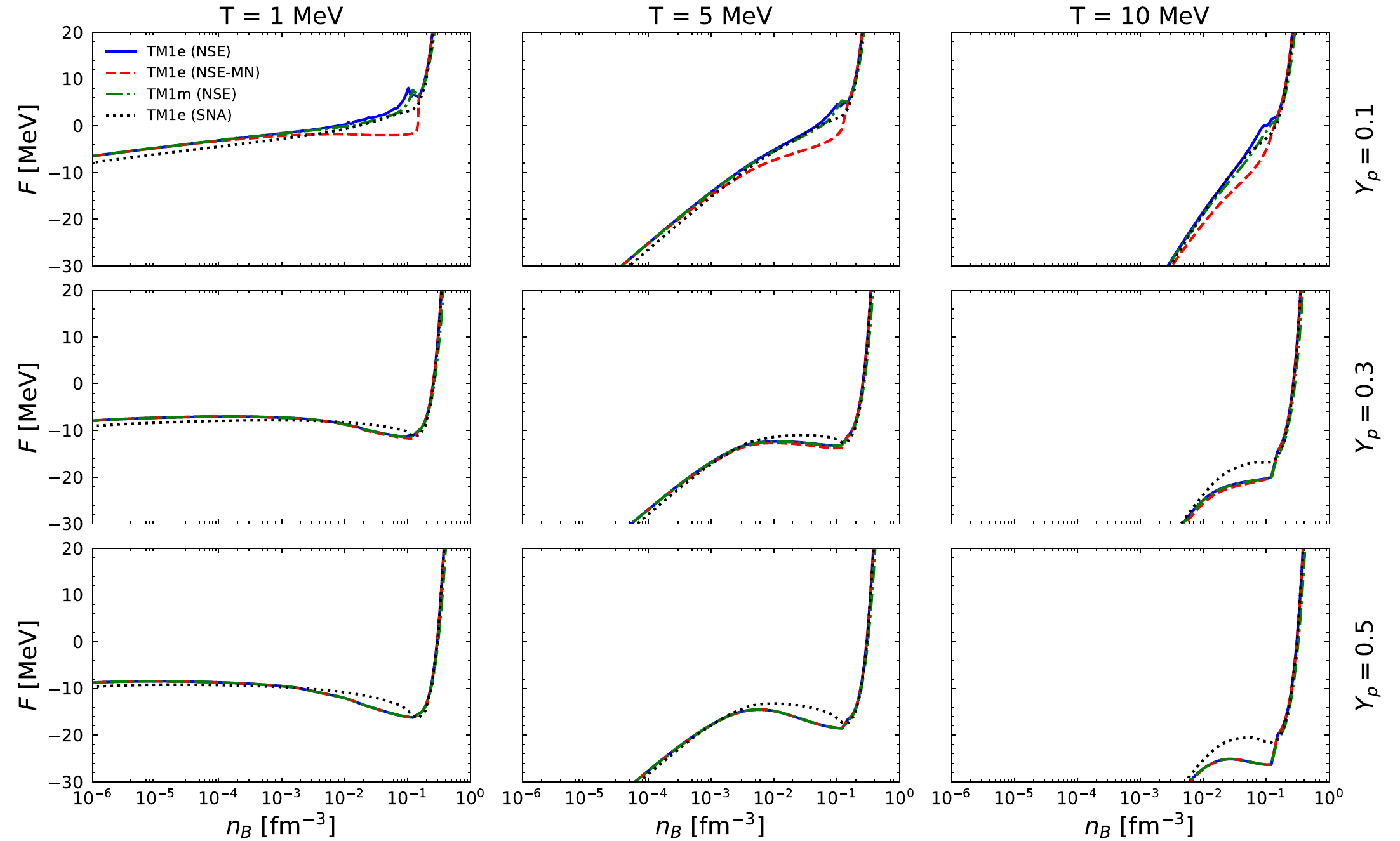}
  \caption{Free energy per baryon $F$.  Line styles are the same as in Fig.~\ref{fig:Mass_Fraction}. }
  \label{fig:Free}
\end{figure*}

\begin{figure*}[htbp]
  \centering
  \includegraphics[width=\linewidth]{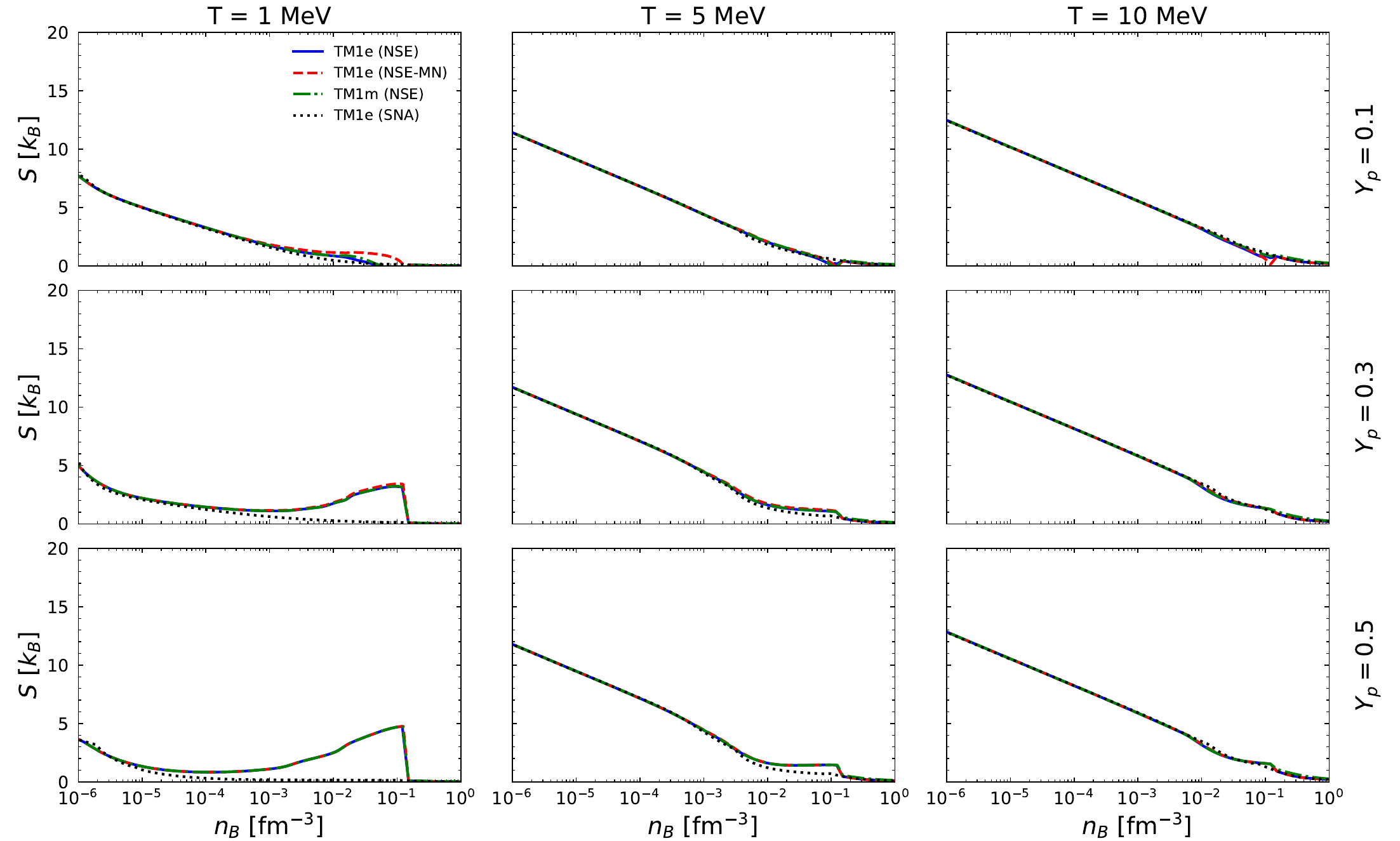}
  \caption{Entropy per baryon $S$. Line styles are the same as in Fig.~\ref{fig:Mass_Fraction}.}
  \label{fig:Entropy}
\end{figure*}

\begin{figure*}[htbp]
  \centering
  \includegraphics[width=\linewidth]{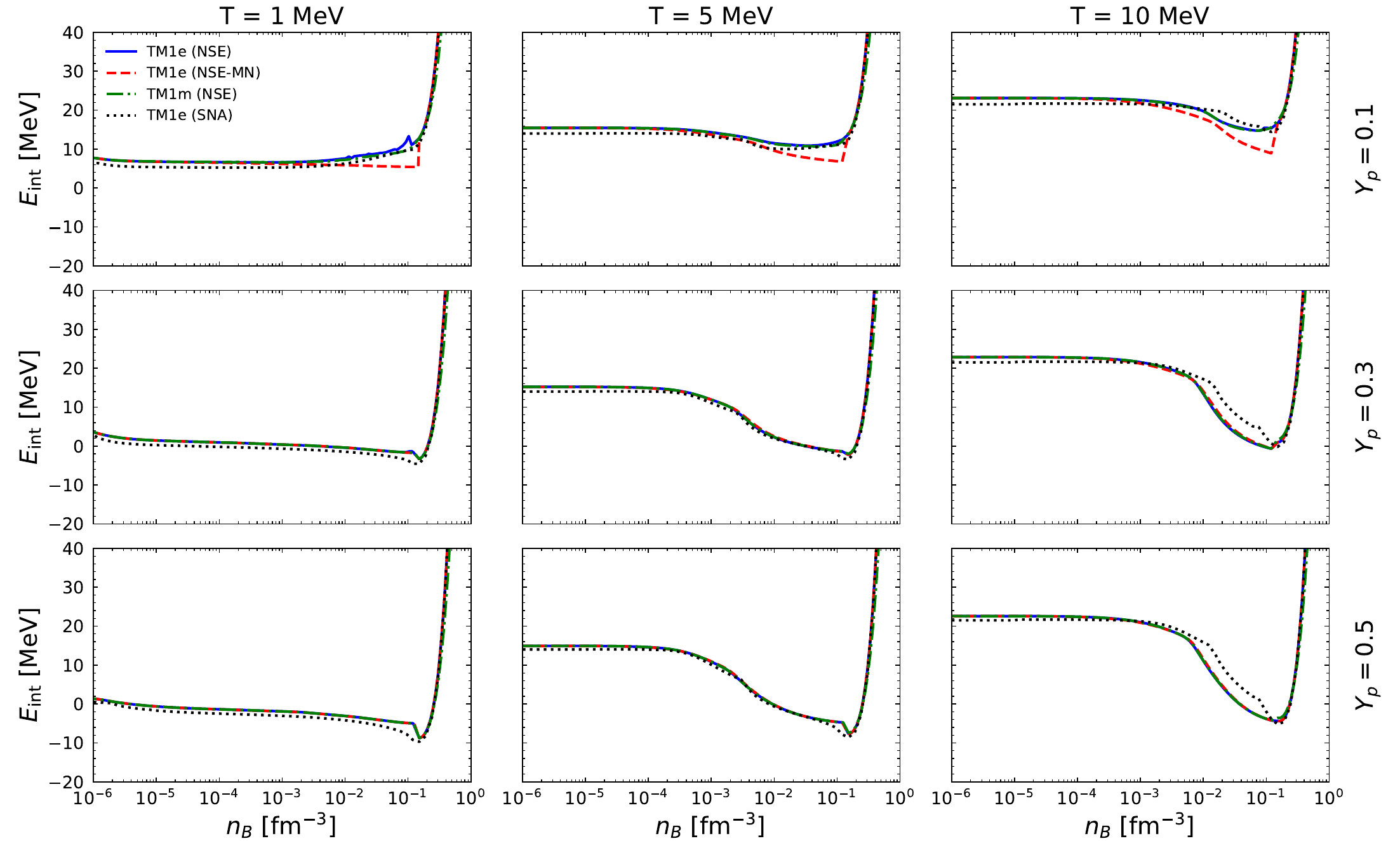}
  \caption{Internal energy per baryon $E_{\rm{int}}$.  Line styles are the same as in Fig.~\ref{fig:Mass_Fraction}.}
  \label{fig:Eint}
\end{figure*}

\begin{figure*}[htbp]
  \centering
  \includegraphics[width=\linewidth]{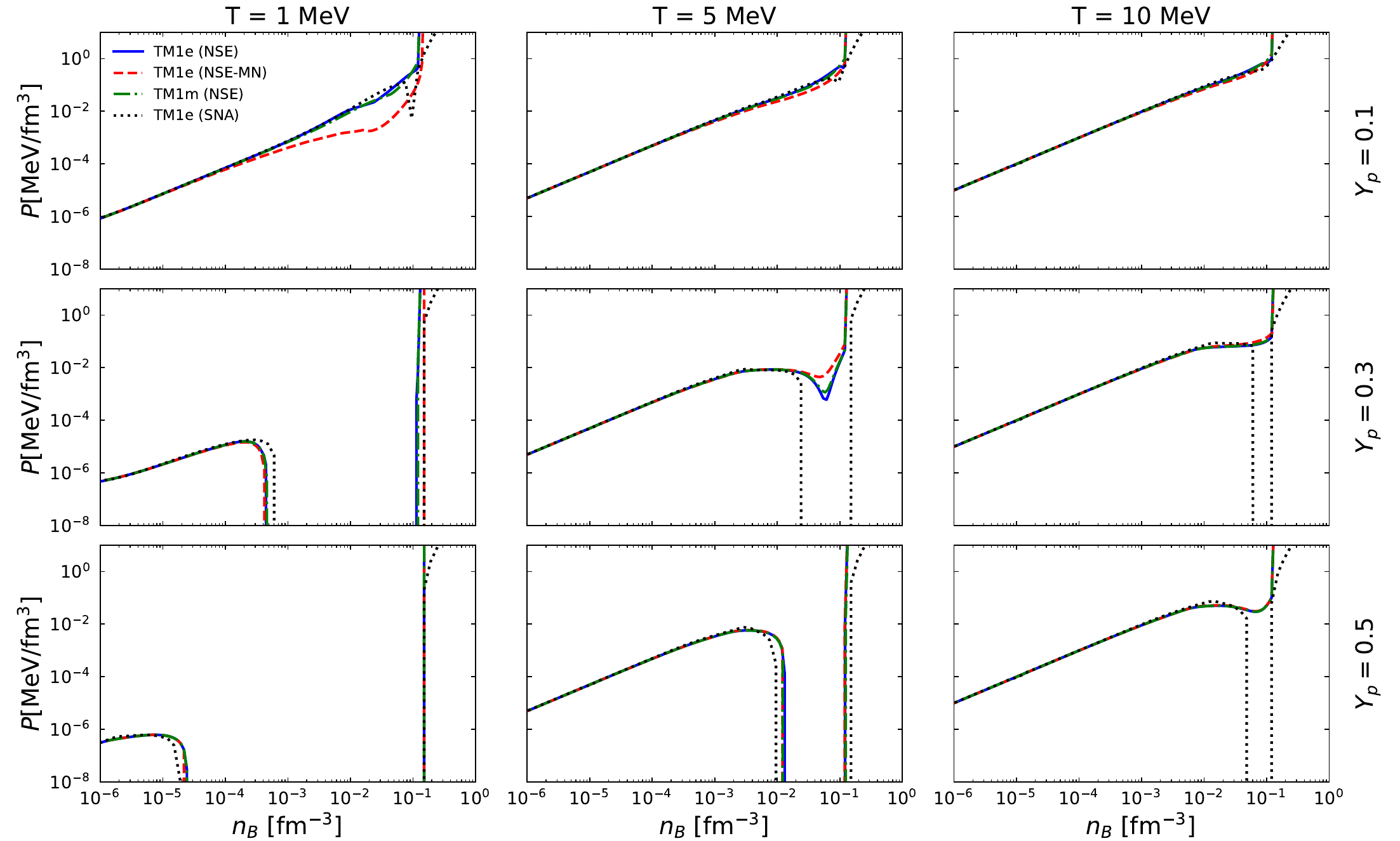}
  \caption{The baryonic pressure $P$.  Line styles are the same as in Fig.~\ref{fig:Mass_Fraction}.}
  \label{fig:Pressure}
\end{figure*}

\newpage

\bibliography{paper}
\end{document}